\documentclass[aps,prl,10pt,twocolumn,showpacs,preprintnumbers,amsmath,amssymb,superscriptaddress]{revtex4-1}

\usepackage{float}
\usepackage{graphicx}
\usepackage{tikz}
\usepackage{dcolumn}
\usepackage{color}
\usepackage{amsmath}

\begin{document}
\title{Nonordinary edge criticaliy of two-dimensional quantum critical magnets}

\author{Lukas Weber}
\affiliation{Institut f\"ur Theoretische Festk\"orperphysik, JARA-FIT and JARA-HPC, RWTH Aachen University, 52056 Aachen, Germany}

\author{Francesco Parisen Toldin}
\affiliation{Institut f\"ur Theoretische Physik und Astrophysik, Universit\"at W\"urzburg, Am Hubland, 
97074 W\"urzburg, Germany}

\author{Stefan Wessel}
\affiliation{Institut f\"ur Theoretische Festk\"orperphysik, JARA-FIT and JARA-HPC, RWTH Aachen University, 52056 Aachen, Germany}

\date{\today}

\begin{abstract}
Based on large-scale quantum Monte Carlo simulations, we examine the 
correlations along  the  edges  of two-dimensional semi-infinite quantum critical Heisenberg spin-$1/2$  systems.
In particular, we consider coupled quantum spin-dimer systems at their  bulk quantum critical points, 
including the columnar-dimer model and the plaquette-square lattice.
The alignment of the edge spins strongly affects these correlations and the
corresponding scaling exponents, with remarkably similar values obtained for various  quantum spin-dimer systems. 
We furthermore observe subtle  effects on the scaling behavior from 
perturbing  the edge spins that exhibit the genuine quantum nature of these edge states. 
Our observations furthermore  challenge recent attempts  that relate
the edge spin criticality  to the presence of symmetry-protected topological phases in such  quantum spin systems.
\end{abstract}

\maketitle

Quantum criticality in quantum many-body systems is a central aspect of current research in condensed matter physics~\cite{Sachdev11}.
In this respect,  quantum  spin systems in particular allow for a detailed comparison of  experimental results to a quantitative computational modeling and analytical calculations of critical properties. 
Prominent examples are  dimerized antiferromagnets,
in which an explicit dimerization of the exchange couplings can be varied (e.g., by applying pressure~\cite{Ruegg04,Ruegg05,Ruegg08,Merchant14}) in order to induce quantum phase transitions between quantum disordered phases and  conventional antiferromagnetic order.
In the absence of frustration,  the quantum critical properties of such systems in $d$ spatial dimensions 
are generally considered  to be in  accord  with the universality class of the $(d+1)$-dimensional  classical Heisenberg model at its finite temperature critical point, described by the Wilson-Fisher fixed point of the three-component $\phi^4$ theory~\cite{ZinnJustin02}. 
This  rationale is  supported  also by large-scale numerical studies of  coupled spin dimer models on various two-dimensional (2D) lattices~\cite{Troyer97,Matsumoto01,Wang06, Wenzel08,Jiang12,Zhang17}. 
Within the nonlinear $\sigma$-model  description of quantum antiferromagnets~\cite{Haldane81,Haldane83a,Haldane83b,Haldane85,Affleck85a,Affleck85b,Chakravarty88,Chakravarty89,Chubukov94},
such an agreement with the  critical  $\phi^4$ theory suggests that for this purpose
spin Berry-phase contributions~\cite{Haldane88} can be neglected in the effective action for 2D dimerized quantum antiferromagnets~\cite{Chubukov94} 
(they may however give rise to additional scaling corrections from cubic terms in coupled dimer systems with reduced spatial symmetries~\cite{Fritz11}). 
As is well known, this   is in stark contrast to the one-dimensional (1D) Heisenberg spin-$1/2$ chain, for which uncompensated spin Berry phases 
lead  to a nonvanishing topological $\theta$-term in the effective continuum action, associated with a
gapless, quantum critical ground state~\cite{Haldane81,Haldane83a,Haldane83b,Affleck85a,Affleck85b,Haldane85}. 
Such a topological term can  also emerge for a one-dimensional edge of 2D quantum spin systems: 
by appropriately cutting  a  2D quantum antiferromagnet to a semi-infinite system, an effective 1D edge spin-$1/2$ system with similarly  uncompensated Berry phases  is generated. 
Such edge spins are furthermore susceptible to  effective interactions induced by the
coupling of the edge spins to the bulk. If the bulk system resides within the quantum disordered region, these effective interactions
along the edge spins  decay exponentially over a length scale set by the finite bulk correlation length. Due 
to the bipartite lattice structure, they 
 respect a bipartite alignment of the edge spin chain, and thereby  lead to long-distance ground state
correlations as in a  spin-$1/2$ Heisenberg chain~(see Supplemental Material~\cite{sm}).  The presence of such  
gapless edge states  of dimerized bulk systems was furthermore  found to be stable against various 
perturbations~\cite{Suzuki12}, whereas the scaling properties  were found to strongly depend on the model parameters and the nature of the applied perturbation~\cite{Suzuki12}.

Here, we   consider    edge spin systems for which  the bulk itself is  tuned onto a quantum
critical point: the long-ranged  quantum critical bulk fluctuations  then dominate the effective interactions 
among the edge spins, which effects changes in  the scaling properties of the correlations  along the edge. 
As for bulk criticality, one may  consider a comparison to  surface critical phenomena in
classical  systems, for which  several scenarios
can be distinguished regarding the bulk vs surface critical behavior~\cite{Binder74, Binder83, Diehl86}. In addition to the ordinary transition, at which the surface
is  critical due to  the bulk transition, the surface may also order at a higher temperature scale than the bulk. 
Such a surface transition  typically requires  enhanced interactions at the surface with respect to those of the bulk, in order to compensate for  the reduced coordination along the surface.
At the bulk transition temperature, the ordered surface may in this case still exhibit additional singular behavior,  known as the  extraordinary transition.
One may furthermore fine-tune the surface coupling  to  a multi-critical special transition,  at which 
surface and bulk are  critical simultaneously. 
Based on the quantum-to-classical correspondence, one would expect the edge spins of a semi-infinite quantum critical spin system to similarly exhibit  genuine quantum critical behavior.
However, it should be noted that a   SU(2)-symmetric 2D quantum system corresponds to a  3D 
classical Heisenberg system with O(3)  symmetry, for which a 2D surface may not order at  finite-temperatures~\cite{Mermin66},
in contrast to the generic scenario outlined above. 
Instead, it has been suggested  that in this case the surface may exhibit a Kosterlitz-Thouless transition  upon varying 
the surface coupling~\cite{Deng05}. 
A direct analogy to the classical case is furthermore exacerbated by the fact that for a 2D quantum spin system 
additional terms to the effective action  from the Berry phases of the edge spins  may affect the critical properties in subtle ways.  Below we provide 
evidence that this is indeed the case. 
%
\begin{figure}[t]
\includegraphics[width=\columnwidth]{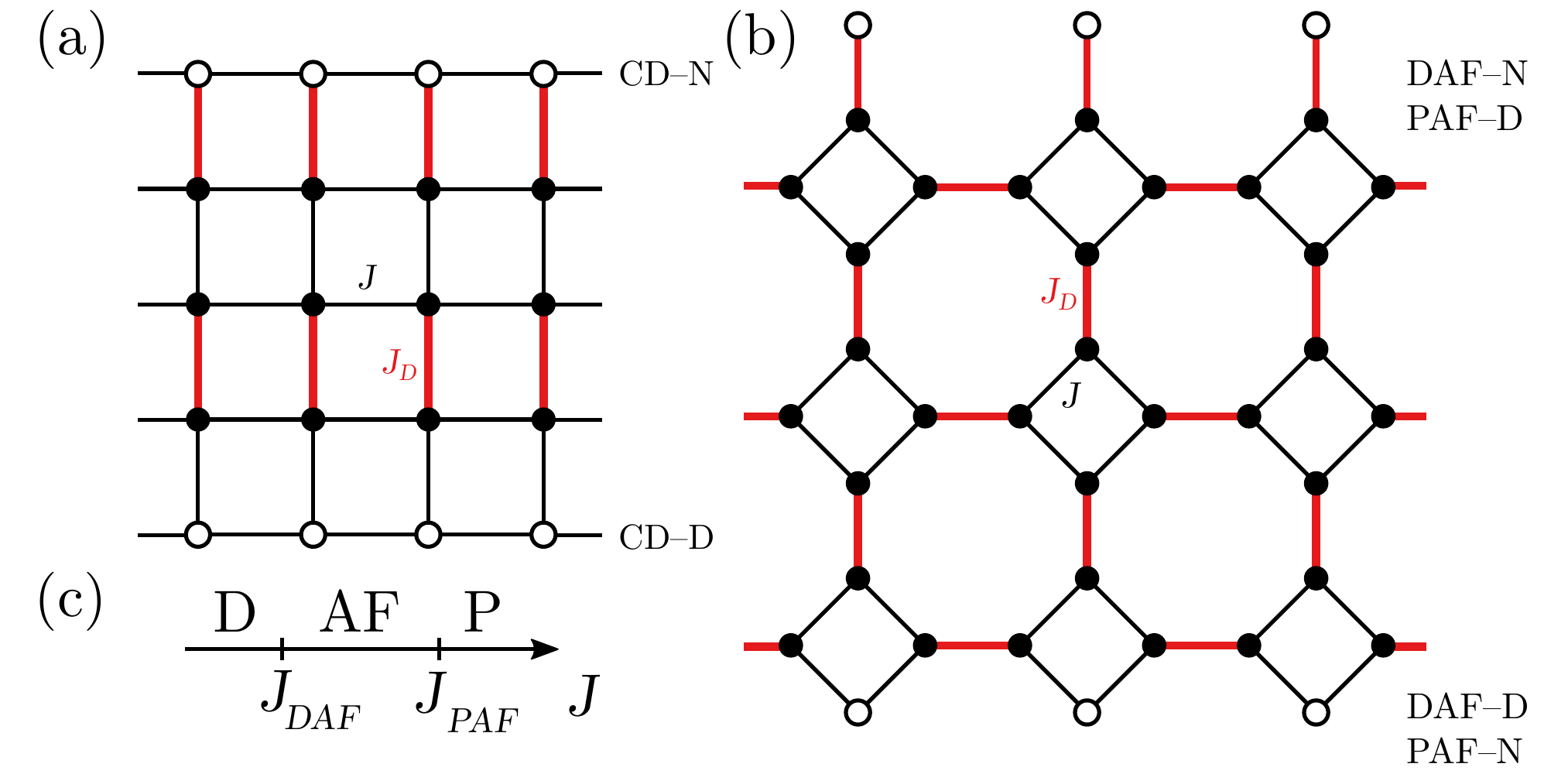}
\caption{(Color online)  (a)  CD lattice with an edge of dangling spins  shown on the bottom edge (CD-D) and nondangling  spins on the top edge (CD-N). (b) PS lattice with dangling (nondangling) edge spins on the bottom (top) edge, for the DAF-D and DAF-N cases, respectively,
and with dangling (nondangling) edge spins on the top (bottom) edge, for the PAF-D and PAF-N cases, respectively.
In both panels,  the interdimer (intra-dimer) couplings $J$ ($J_D$) are indicated  by black (bold red) lines,  the edge (bulk) spins by open (full) symbols, and periodic boundary conditions by open lines. 
(c) Phase diagram of the PS lattice with the antiferromagnetic region (AF), the dimer phase (D) for $J<J_{DAF}\approx 0.6J_D$, and the plaquette phase (P) for $J>J_{PAF}\approx1.1J_D$.
}
\label{fig_lattices}
\end{figure}
%
For this purpose, we  examined several specific coupled spin-dimer systems that are described by a generic Hamiltonian
$
H=J \sum_{\langle i,j \rangle} \mathbf{S}_i \cdot  \mathbf{S}_j +  J_D \sum_{\langle i,j \rangle_D} \mathbf{S}_i \cdot  \mathbf{S}_j ,
$
where the first (second) term contains the interdimer (intradimer) couplings of strength $J$ ($J_D$) of different geometries, to be specified  below
(we fix $J_D=1$). 
We performed  quantum Monte Carlo  (QMC) simulations of such coupled spin-dimer systems using the stochastic series expansion~\cite{Sandvik91} approach with 
deterministic operator loop updates~\cite{Sandvik99, Henelius00}. For a 2D system, the number of spins scales as $L^2$ with the linear system size, and in order to probe ground state properties
we scaled the temperature as $T=1/(2L)$, respecting  the dynamical critical exponent $z=1$ for the bulk transition.
In the following, we present the QMC results for various
edge configurations. 

We first  consider the columnar-dimer  (CD) lattice  shown in Fig.~\ref{fig_lattices}(a). 
Its bulk quantum critical point has been located previously at $J=0.52337(3)$~\cite{Matsumoto01,Wenzel08}. 
Using periodic boundary conditions (PBC) along the lattice direction parallel to the dimers, we  examine separately the two cases of cutting along the perpendicular direction, 
obtaining either an edge of dangling spins (with respect to the $J_D$ bonds), denoted  CD-D, or an edge of nondangling spins, denoted CD-N cf. Fig.~\ref{fig_lattices}(a).
For both cases, 
we  performed QMC simulations to measure the spin-spin 
correlations  $\langle S^z_i S^z_j \rangle$  among  two  edge spins $i,j$ at a distance $r$ parallel to  the edge, denoted $C_\parallel(r)$, as well as 
between an edge spin $i$  and an equivalent bulk spin $j$ (with respect to the unit cell) at a distance $r$ perpendicular to the edge, denoted  $C_\perp(r)$.
In addition, we also accessed
the staggered susceptibility $\chi_s$ of the edge spin subsystem from the Kubo integral~\cite{Sandvik91}, $\chi_s=\frac{1}{L}\int_0^\beta d\tau \: \langle M_s(\tau) M_s(0)\rangle$, of the staggered edge moment $M_s=\sum'_i \varepsilon_i S^z_i$, where the summation is restricted over the edge spins ($\epsilon_i=\pm 1$ depending on the sublattice to which site $i$ belongs). 
%
\begin{figure}[t]
\includegraphics[width=\columnwidth]{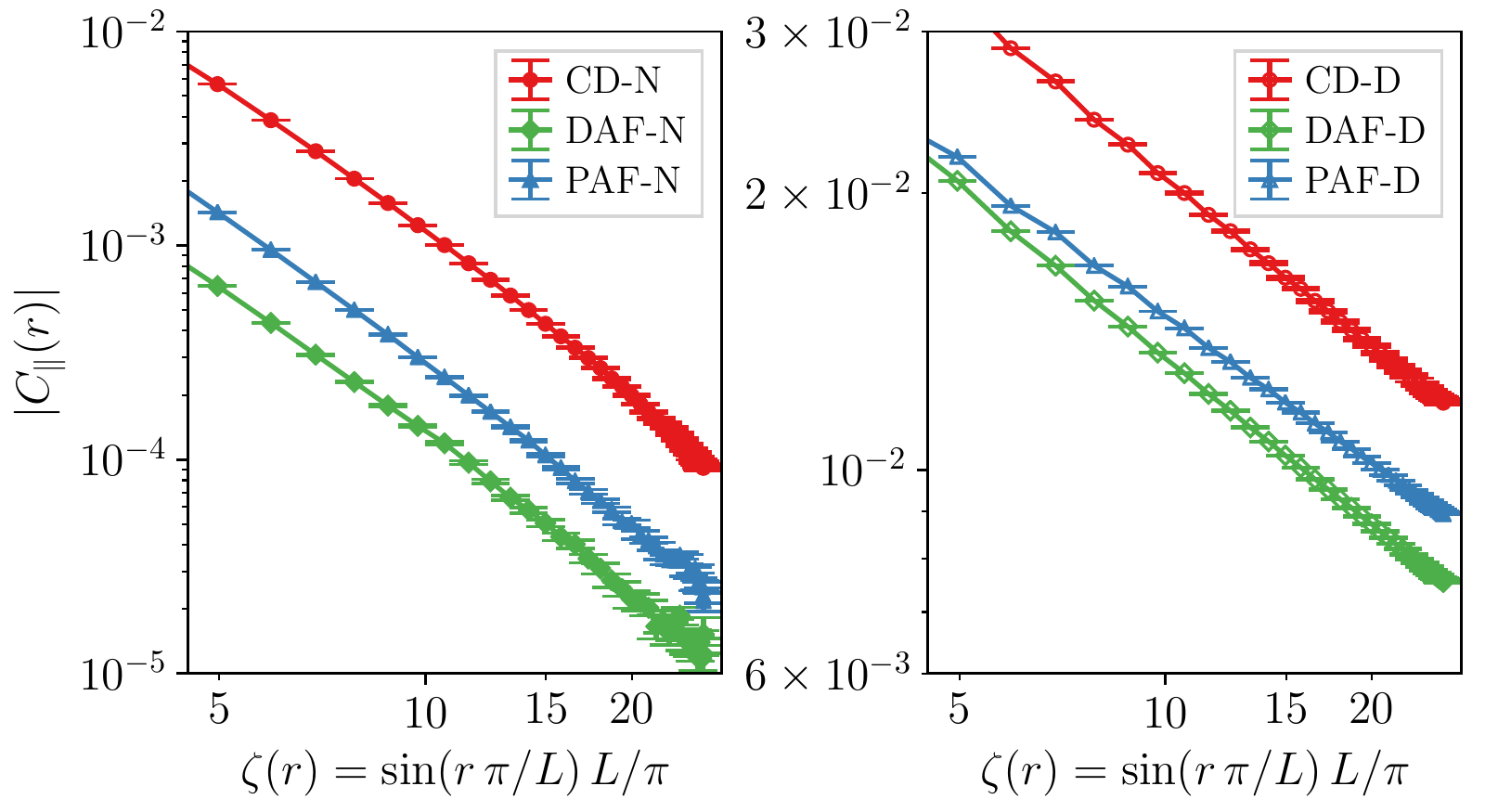} 
\caption{(Color online)  $C_\parallel(r)$  vs. $\zeta(r)=\sin(r\pi/L)L/\pi$ for the different edge spin configurations shown in Fig.~\ref{fig_lattices}.
}
\label{fig_Cparallel}
\end{figure}
%
The resulting data for  $C_\parallel(r)$ on a $L=80$ system  is shown (along with that for several other cases, discussed below) in Fig.~\ref{fig_Cparallel}. It  shows
$C_\parallel(r)$
as a function of the conformal length (cord distance) $\zeta(r)=\sin(r\pi/L)L/\pi$, to account for the PBC along the edge. 
For both cases, we observe an approximately algebraic decay, indicative of a  quantum critical state of the
edge spin system that can be quantified by the scaling  
$
|C_\parallel(r)|\propto r^{-z-\eta_\parallel}, 
$ 
with an anomalous critical exponent $\eta_\parallel$ and with $z=1$, here and in the following. 
The drop of the correlation functions at large values of $\zeta(r)$, explicitly seen in the weaker-correlated nondangling case,  indicates residual  finite-size effects~(see Supplemental Material~\cite{sm}). 
To account  for  finite-size corrections, we  estimate $\eta_\parallel$ from
the finite-size scaling of $C_\parallel(L/2)$ vs $L$ as $C_\parallel(L/2)=(L/2)^{-z-\eta_\parallel}(c_0+c_1 L^{-\omega}) $,
including a subleading scaling correction (in practice, we fix $\omega=1$~\cite{Zhang17,sm}), and $c_0$ and $c_1$ as nonuniversal fit parameters.  
We  obtain this way the  estimates 
$\eta_\parallel=-0.50(1)$ (CD-D) and  $\eta_\parallel=1.30(2)$ (CD-N), respectively, cf. Table~\ref{table1} and the scaling plots in the top panel of Fig.~\ref{fig_CparallelLo2}.
\begin{figure}[t]
\includegraphics[width=\columnwidth]{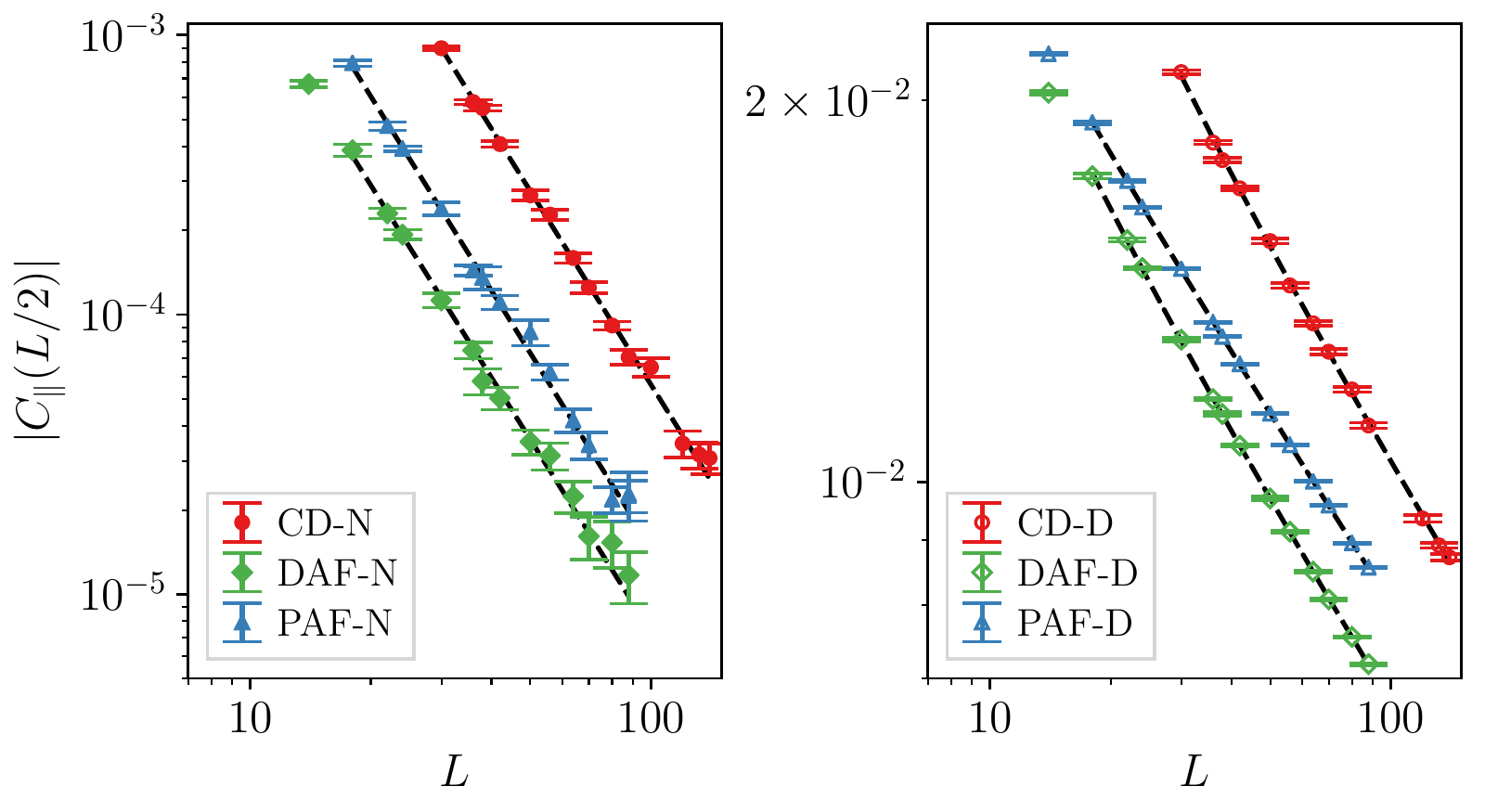} 
\includegraphics[width=\columnwidth]{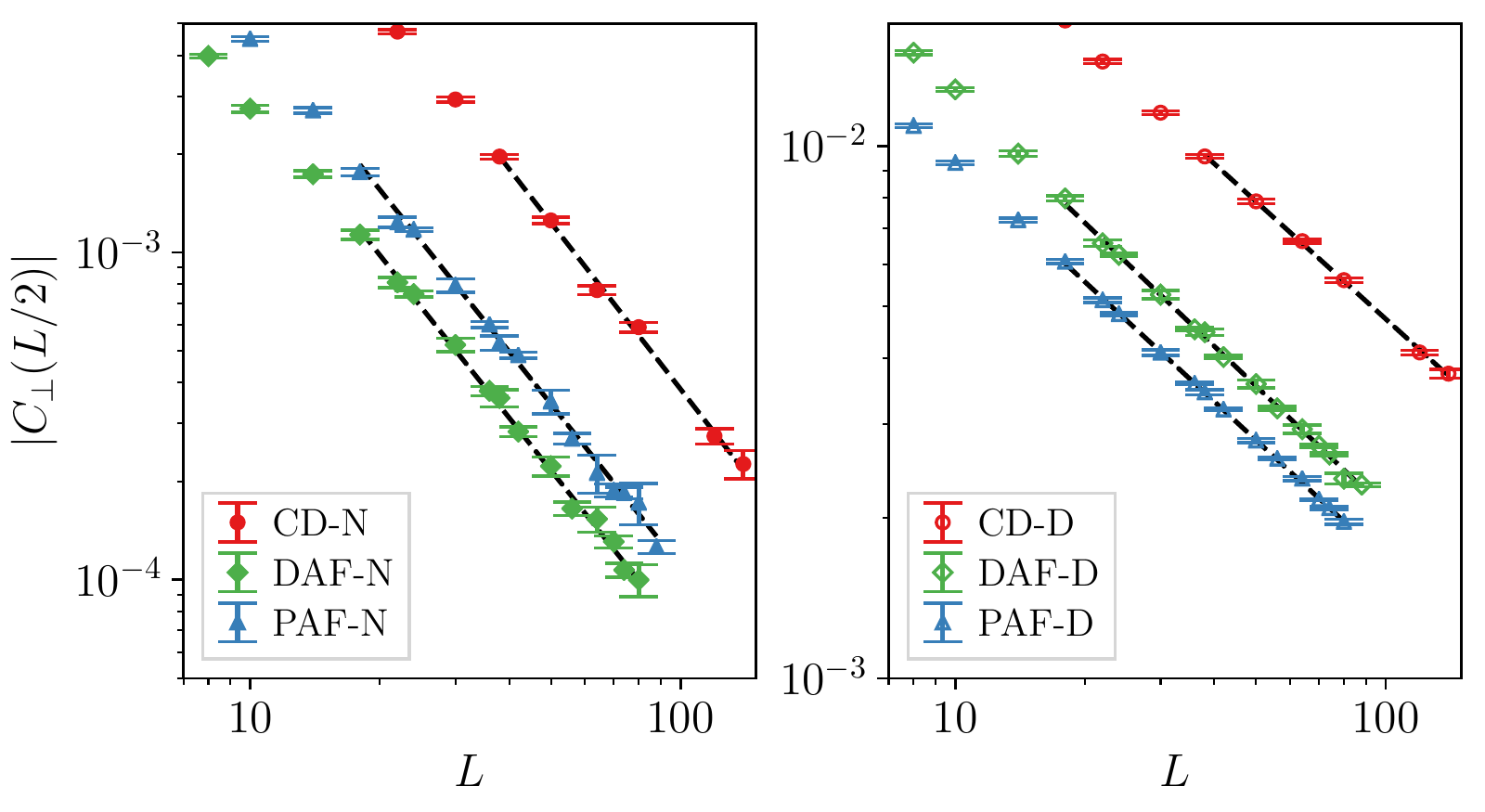} 
\includegraphics[width=\columnwidth]{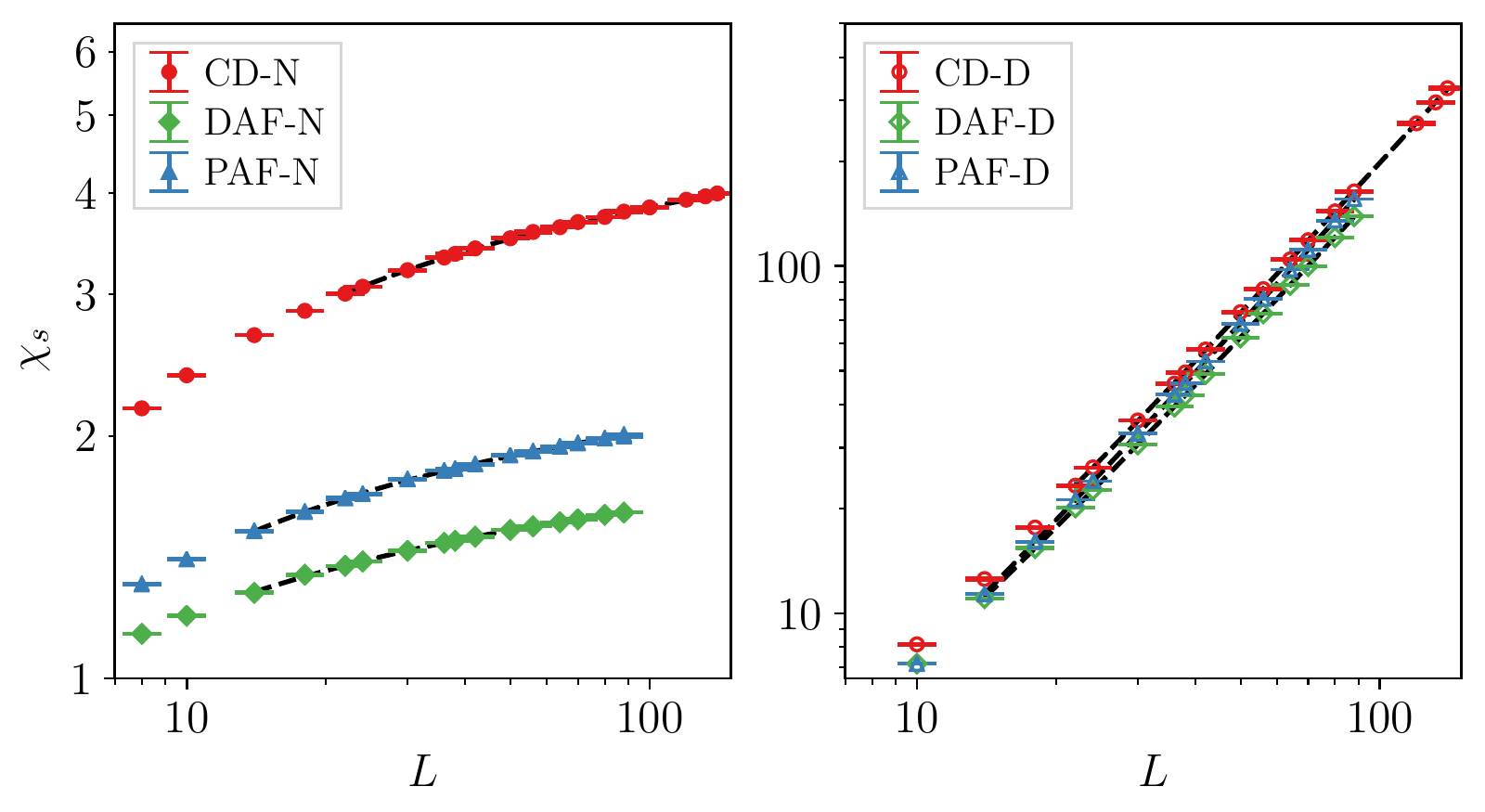} 
\caption{$C_\parallel(L/2)$ (top panels),  $C_\perp(L/2)$  (middle panels), and $\chi_s$  (bottom panels)  as  functions of  $L$ for the different  configurations of Fig.~1 along with finite-size fits (dashed lines).
}
\label{fig_CparallelLo2}
\end{figure}
We  also observe 
scaling for $C_\perp(L/2)$ (cf. Fig.~\ref{fig_CparallelLo2}), and from a corresponding fit  to $C_\perp(L/2)=(L/2)^{-z-\eta_\perp}(c_0+c_1 L^{-\omega})$ obtain 
the estimates for 
$\eta_\perp$ provided in  Table~\ref{table1}~(see Supplemental Material~\cite{sm}).
Furthermore,  from the finite-size scaling $\chi_s\sim L^{-(1+z-2y_{h_1})}$, we estimate the scaling dimension $y_{h_1}$ of the (staggered) field along the edge.
With $z=1$,
this  scaling   corresponds to the standard form for classical surface critical behavior~\cite{Binder74, Binder83, Diehl86} for a 3D bulk system in terms of the surface field scaling dimension $y_{h_1}$,
and for which  the following scaling relations hold:
$
\eta_\parallel=3-2y_{h_1}, \quad 2\eta_\perp=\eta_\parallel+\eta.
$
Here,  $\eta$ is the anomalous dimension at the bulk transition, with
$\eta=0.0375(5)$ for the 3D O(3) universality class~\cite{Campostrini02}.
We can estimate $y_{h_1}$ from the QMC data based on the finite-size scaling form
$\chi_s= c_{ns} + L^{-(1+z-2y_{h_1})}(c_0+c_1 L^{-\omega})$ that
includes an additive constant $c_{ns}$ to account for regular  contributions in the nondangling case~(see Supplemental Material~\cite{sm}).
The obtained estimates for $y_{h_1}$ are  listed in  Table~\ref{table1} and the scaling plots are shown in Fig.~\ref{fig_CparallelLo2}.
\begingroup
\squeezetable
\begin{table}
\begin{ruledtabular}
\begin{tabular}{l || c | c | c}
Configuration & $\eta_\parallel$ & $\eta_\perp$ & $y_{h_1 }$ \\
\hline\hline
CD-N & 1.30(2) & 0.69(4) & 0.84(1)\\
DAF-N & 1.29(6) & 0.65(3) & 0.832(8)\\
PAF-N & 1.33(4) & 0.65(2) & 0.82(2)\\
\hline
CD-D & -0.50(1) & -0.27(1) & 1.740(4)\\
DAF-D & -0.50(1) & -0.228(5) & 1.728(2)\\
PAF-D & -0.517(4) & -0.252(5) & 1.742(1)\\
\end{tabular}
\end{ruledtabular}
\caption{Critical exponents $\eta_\parallel$,  $\eta_\perp$, and $y_{h_1 }$ for the edge spin configurations of 2D coupled spin-dimer  systems in Fig.~\ref{fig_lattices}.
}
\label{table1}
\end{table}
\endgroup
For both edge spin configurations the critical exponents obey the above  scaling relations   
to the precision of  their estimated uncertainties. 
%
 
We next consider the plaquette-square (PS) lattice~\cite{Troyer96,Troyer97}; cf. Fig.~\ref{fig_lattices}(b). This model has been analyzed in the context of edge  spin criticality 
in a recent publication~\cite{Zhang17}, and we comment on the conclusions drawn by this work further below. 
Here, we  consider PBC in the horizontal and open boundary conditions in the vertical direction; cf.~Fig.~\ref{fig_lattices}(b). As a function of the coupling ratio $J/J_D$, this system shows two quantum critical
points, at $J=J_{DAF}=0.603520(10)J_D$ and for $J=J_{PAF}=1.064382(13)J_D$. They  separate the antiferromagnetic  phase obtained for $J\approx J_D$ from the quantum-disordered dimer-singlet  (plaquette-singlet) dominated phase for $J<J_{DAF}$  ($J>J_{PAF}$), respectively (we consider $J,J_D>0$). Noting  the difference between the two quantum-disordered phases with respect to the pattern of the predominant singlet formation, we   distinguish the following four different edge spin configurations.
(i) For $J<J_{DAF}$,  the system is quantum disordered due to  predominant singlet formation along the $J_D$ dimer bonds,
and  thus the systems  exhibits dangling spins if we cut through a row of dimers to obtain the bottom boundary in  Fig.~\ref{fig_lattices}(b). At $J=J_{DAF}$ , we hence denote this edge spin configuration as DAF-D. 
(ii) If for $J<J_{DAF}$ we instead consider the  spins at  the top edge in Fig.~\ref{fig_lattices}(b), we obtain nondangling spins and for  $J=J_{DAF}$ we denote this edge spin configuration as DAF-N.
(iii)  For $J>J_{PAF}$,  the system is instead quantum disordered due to   predominant  four-site singlet formation on the plaquettes formed  by the $J$ bonds, 
and   the system thus exhibits dangling spins at the top edge in  Fig.~\ref{fig_lattices}(b). At $J=J_{PAF}$, we thus denote this edge spin configuration as PAF-D. 
(iv) If for $J>J_{PAF}$  we instead consider  spins at  the bottom edge in Fig.~\ref{fig_lattices}(b), we obtain nondangling spins and for  $J=J_{PAF}$ we denote this edge spin configuration as PAF-N.
For the  PS lattice, we can thus realize both the case of dangling and the nondangling edge  spins at either quantum critical point by considering  appropriate edges. The  QMC data for the correlation function
$C_\parallel(r)$ 
for both cases are also shown in 
Fig.~\ref{fig_Cparallel}. Performing again a finite-size scaling analysis of the correlation functions $C_\parallel(L/2)$ and $C_\perp(L/2)$
as well as the staggered susceptibility $\chi_s$ of the edge spins~(see Supplemental Material~\cite{sm}),  we  obtain scaling exponents that  essentially correspond to   those for the other considered cases; cf.~Table~\ref{table1} and Fig.~\ref{fig_CparallelLo2}.

At the considered quantum critical points,  which all belong to the 3D O(3) universality class, the edge spins exhibit critical scaling exponents that apparently belong to 
two different  classes, depending on whether the edge spins are dangling or not with respect to the predominant singlet formation in the neighboring quantum disordered phase [we  find consistent exponents also for the square-lattice bilayer model with dangling (nondangling) edge spins~(see Supplemental Material~\cite{sm})].
For the nondangling case, the obtained critical
exponent $y_{h_1}$ is similar to  the values $y_{h_1}=0.813(2)$ and $y_{h_1}=0.802(1)$  obtained from Monte Carlo~\cite{Deng05} and conformal bootstrap~\cite{Gliozzi15} studies of the ordinary surface transition in the 3D O(3) model, respectively. This is in accord with  the expectation that  the critical behavior for  nondangling edge spins is induced by the quantum critical fluctuations of the bulk system.
The estimated exponents are  comparable even  to the values $\eta_\parallel=1.307$, $\eta_\perp=0.664$ and $y_{h_1}=0.846$, obtained for the ordinary surface transition from the second-order $\epsilon$ expansion of the O($n$)-symmetric vector model ($\epsilon=4-d$)~\cite{Diehl81,Diehl86}, after a d\'egag\'e evaluation at  $\epsilon=1$ and $n=3$~\cite{Ding18}. 
Regarding the dangling case, one observes a  similar closeness of the critical exponents to the  values $\eta_\parallel=-0.445$, $\eta_\perp=-0.212$ and $y_{h_1}=1.723$ obtained from  second-order  $\epsilon$ expansion
 for the special  transition~\cite{Diehl81,Diehl86}, evaluated at $\epsilon=1$ and $n=3$~\cite{Ding18}; however, there is still some spread among the values in Table~\ref{table1} and these estimates~\cite{footnote2}. 
Moreover, as  mentioned above, the 3D  O(3) model does not feature such a special transition, whereas the $\epsilon$ expansion is blind to this restriction~\cite{Diehl86}.  To assess if this apparent similarity of the 
critical exponents extends beyond a mere coincidence  or if  fine-tuning is necessary, requires, e.g.,   an  $\epsilon$ expansion in  the
presence of a  $\theta$ term from the dangling edge spins, to be compared to 
the $\epsilon$ expansion for the classical special transition,  evaluated at $n=3$. We are not aware of such an argument.
 
In Ref.~\cite{Zhang17}, the observation that the scaling exponents  for the DAF-D configuration differ from the ordinary transition is argued to be   a consequence of 
symmetry-protected-topological (SPT) order~\cite{Gu09,Chen12} in the  ground state for $J<J_{DAF}$ in the form of an Affleck-Kennedy-Lieb-Tasaki state~\cite{Affleck87}---in contrast to the
trivial (non-SPT) nature of, e.g., the
plaquette phase or the quantum-disordered phase of the CD model. 
However, we  obtain such nonordinary exponents also  in the PAF-D configuration at  $J=J_{PAF}$ as well as for the critical CD model with dangling spins. 
The  nonordinary edge criticality is thus not a characteristic feature of SPT phases but results from the dangling edge spin arrangement. 
Moreover, it is readily seen to be possible to adiabatically connect the quantum-disordered phase
of the PS model for $J<J_{DAF}$  to the quantum-disordered regime of the
bilayer square lattice model without breaking any 
symmetries of the  PS model~(see Supplemental Material~\cite{sm}). 

\begin{figure}[t]
\includegraphics[width=\columnwidth]{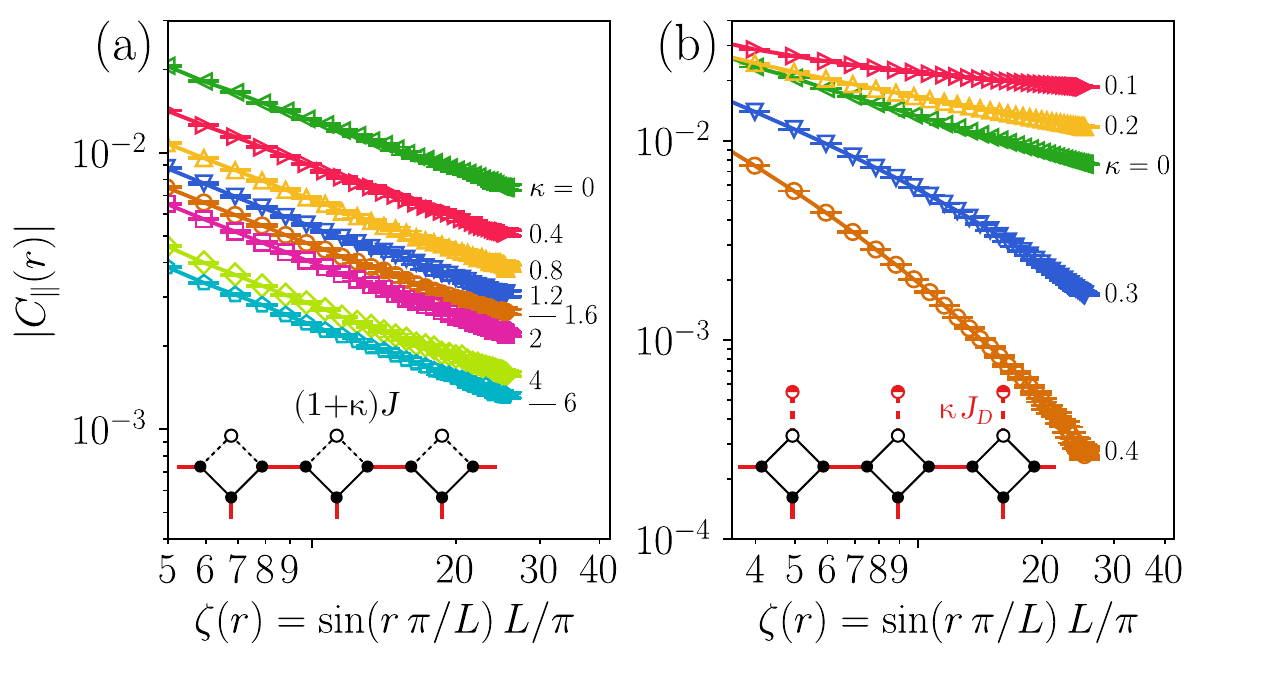} 

\caption{(Color online)
Lateral  correlations   $C_\parallel(r)$ vs. $\zeta(r)$ for the DAF-D configuration with different values of the edge coupling enhancement $\kappa$, as shown in the inset [left panel (a)], and
for the DAF-D configuration with different values of the coupling  $\kappa J_D$ to  additional spins, shown in the inset by semifilled circles [right panel (b)]. In both panels the curves are labeled by the value of $\kappa$.}
\label{fig_added_kappa}
\end{figure}

In order to probe  the stability of the scaling exponents with respect to variations of the edge spin couplings, we  introduced  modifications to  the local environment of the dangling edge spins. We find that depending on 
the specific setting, different scenarios are realized. For example, enhancing  the exchange couplings along the edge in the DAF-D configuration by a relative factor $\kappa$  does apparently not significantly alter the critical properties of the edge spins; cf. Fig.~\ref{fig_added_kappa}(a). Its main effect is a uniform overall reduction of the correlations,
such as if the increased couplings  quench the magnetic moments on 
the edge sites by  forming effective spin-1/2 moments on the $\wedge$-shaped outer triangles. 
On the other hand,  coupling  each dangling spin in the DAF-D configuration to an additional spin, as shown in the inset of   Fig.~\ref{fig_added_kappa}(b), strongly affects the scaling of the original edge spin correlations:  while for small values of the additional coupling $\kappa J_D$, the edge spin correlations increase, they are eventually  suppressed at large values of $\kappa$, as shown in Fig.~\ref{fig_added_kappa}(b). This  nonmonotonous behavior can be understood as follows: The additional couplings initially enhance  antiferromagnetic tendencies, so that weak values of $\kappa$ lead to more extended correlations. On the other hand, a further increase of $\kappa$ 
leads to a  predominant formation  of local singlets on the new bonds, which eventually suppress the long-distance correlations. This perturbation, which allows us to tune from the DAF-D configuration ($\kappa=0$) to the DAF-N configuration ($\kappa=1$), thus exhibits explicitly the genuine quantum nature of these critical edge states.
It may also be intriguing to  examine the possibility of a true phase transition within the edge states along this line.
%
%

Based on the above findings, an analytical approach to the edge spin correlations in quantum critical bulk systems would be desirable, 
in particular in order to  rationalize the apparent similarity of the  exponents for the case of dangling spins with a 
naive extrapolation of the $\epsilon$ expansion for the special surface transition.

\textit{Note added.} Recently, we became aware of a related study~\cite{Ding18}, where consistent numerical findings for the CD model are reported.\\


We thank  F. F. Assaad, A. M. L\"auchli,  and T. C. Lang for useful discussions.
F.P.T. is supported by the Deutsche Forschungsgemeinschaft (DFG) through Grant No. AS/120/13-1 of the FOR1807.
SW and LW acknowledge support by the DFG through Grant No.  WE/3649/4-2 of the FOR 1807 and through RTG 1995.
Furthermore, we thank the IT Center at RWTH Aachen University and the JSC J\"ulich for access to computing time through JARA-HPC.

\newpage
\clearpage
\section{Supplemental Material}

\setcounter{figure}{0}  
\renewcommand{\thefigure}{S\arabic{figure}}
\renewcommand{\thetable}{S\Roman{table}}
\subsection{Edge correlations for a quantum-disordered bulk}

\begin{figure}[H]
\includegraphics[width=\columnwidth]{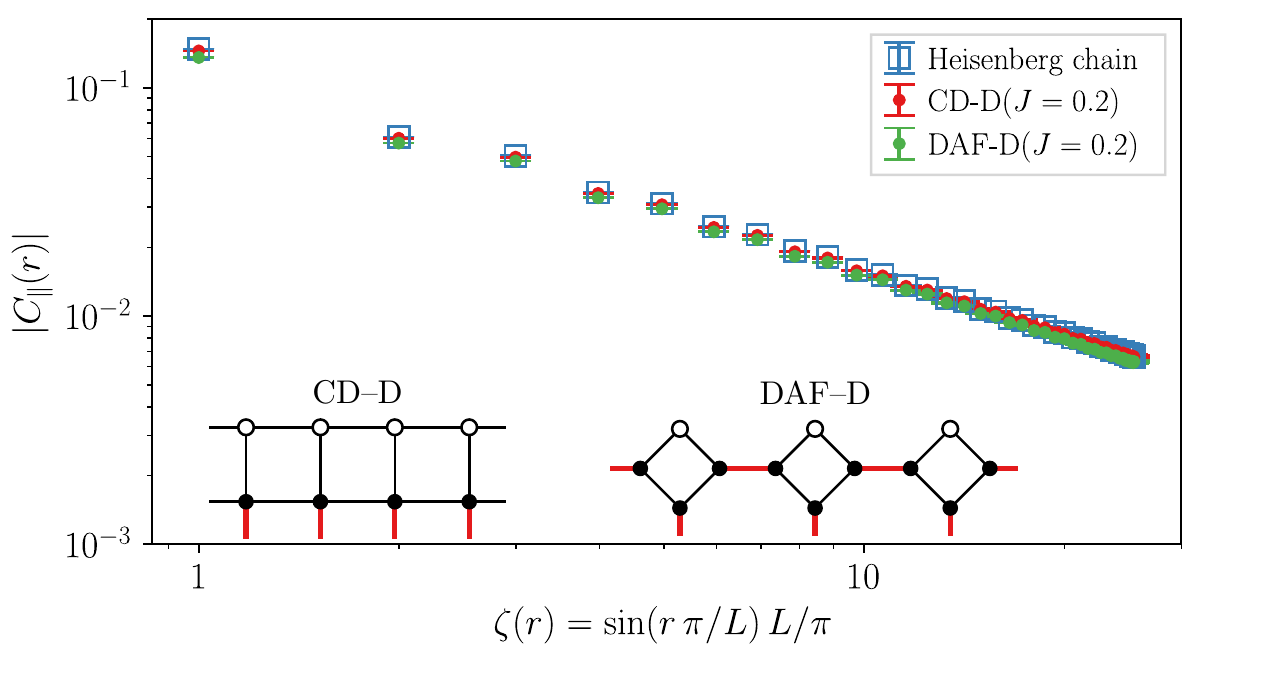} 
\caption{Lateral correlations $C_\parallel(r)$ (circles) as a function of  the cord distance $\zeta(r)$ between the edge spins for 
the two edge spin configurations shown in the insets:  the
dangling edge spin configuration CD-D of the columnar dimer lattice (left inset), 
and  the  configuration dangling edge spin DAF-D of the plaquette-square lattice (right  inset),
for a coupling ratio of $J/J_D=0.2$, based on simulations with
with $L=80$, at a temperature of $T=0.00125J_D$, and  $T=0.0003125J_D$, respectively. 
For comparison, the results of a spin-$1/2$ Heisenberg chain with $80$ sites is also shown (squares). 
}
\label{fig_noncritical}
\end{figure}

As an example of the effective spin-$1/2$ Heisenberg chain-like  correlations  that emerge at the edge of a quantum disordered bulk system, we show in Fig.~\ref{fig_noncritical}
the lateral correlations $C_\parallel(r)$  (circles) as a function of  the distance $r$ between the edge spins for 
 (i) the dangling edge spin configuration CD-D of the columnar dimer lattice (cf. the left inset), 
and (ii) the dangling edge spin configuration DAF-D of the plaquette-square lattice (cf. the right inset)
for a coupling ratio of $J/J_D=0.2$.
For comparison, the corresponding correlation function of a spin-$1/2$ Heisenberg chain 
is also shown in this figure. 

\newpage

\subsection{Finite-size effects in $C_\parallel(r)$}

\begin{figure}[H]
\includegraphics[width=\columnwidth]{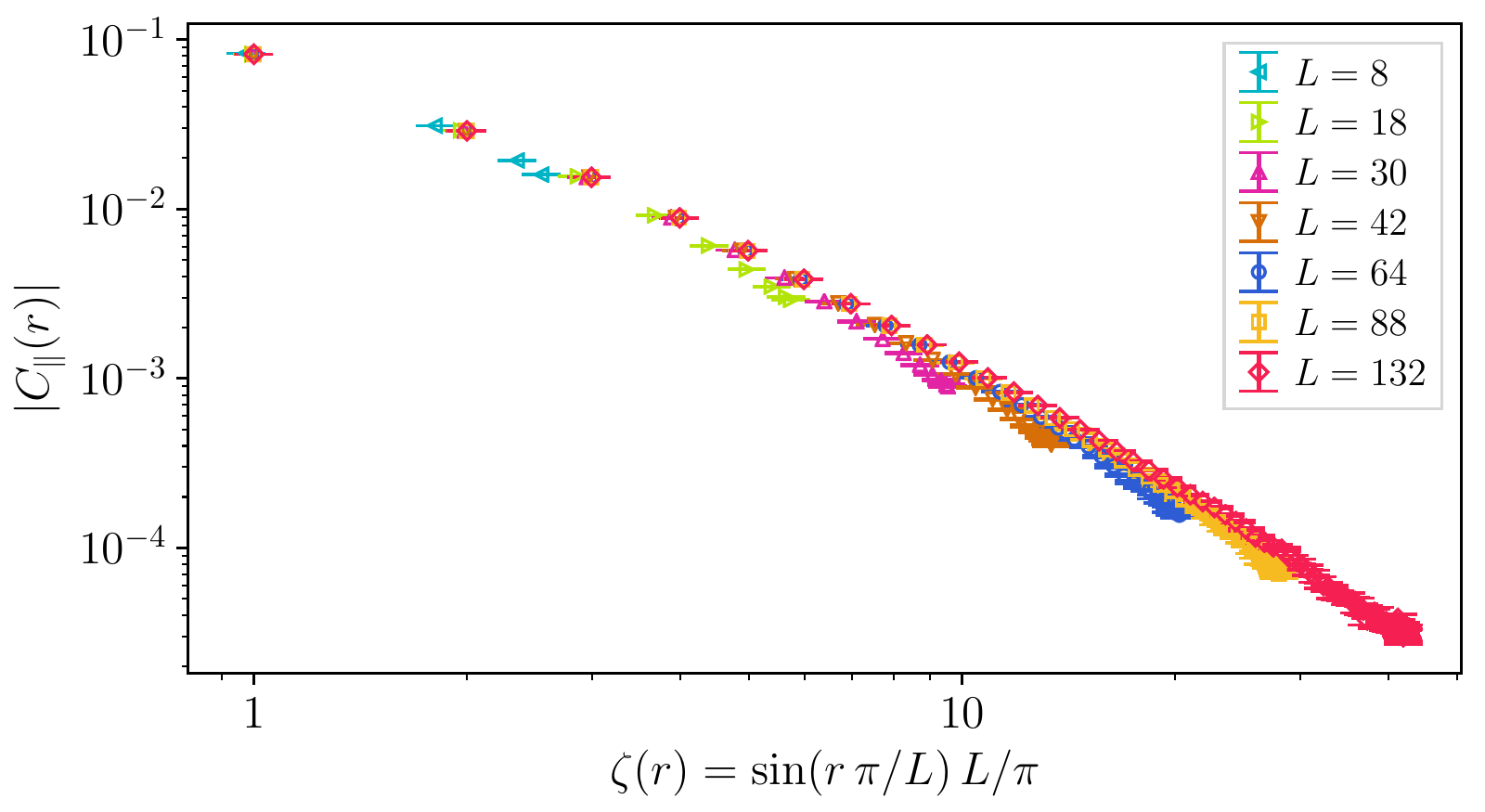} 
\caption{Lateral  correlations   $C_\parallel(r)$  as a function of the conformal distance $\zeta(r)$ along the  edge spins of the CD-N configuration  
 shown in Fig.~1 of the main text for different values of the linear system size $L$. The temperature is scaled as $T=J_D/(2L)$ in all cases. 
}
\label{fig_fs}
\end{figure}

Figure~\ref{fig_fs} shows the lateral correlations   $C_\parallel(r)$  as a function of the conformal distance $\zeta(r)$ along the  edge spins of the CD-N configuration shown in Fig.~1 of the main text for different values of the linear system size $L$.

\newpage
\subsection{Results for the bilayer square lattice}

\begin{figure}[b]
\begin{center}
\includegraphics[width=\columnwidth]{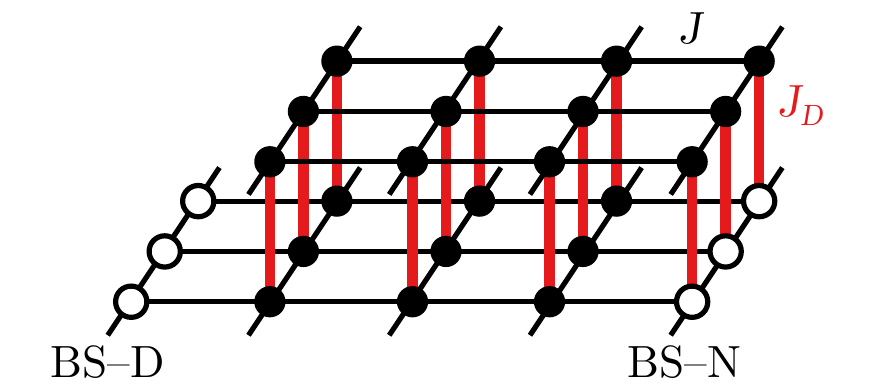} 
\end{center}
\caption{BS lattice with dangling (BS-D) and nondangling (BS-N) edge spins. The $J_D$ bonds (bold-red) here connect the two square lattice layers while the $J$ bonds (black) form the intra-layer couplings. Open (full) circles denote the edge (bulk) spins.}
\label{fig_bilayer}
\end{figure}

\begin{figure}[H]
\begin{center}
\includegraphics[width=\columnwidth]{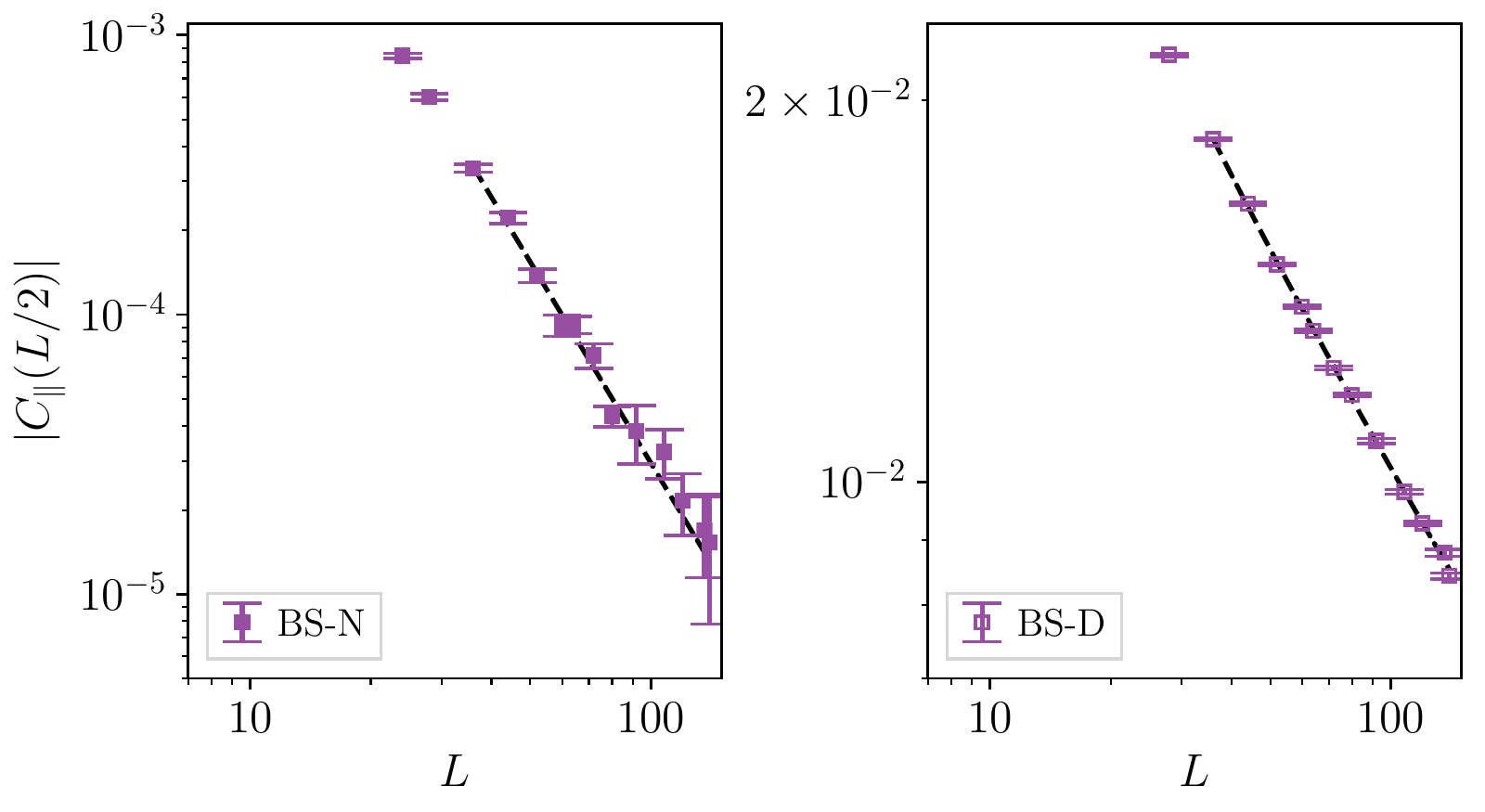} 
\includegraphics[width=\columnwidth]{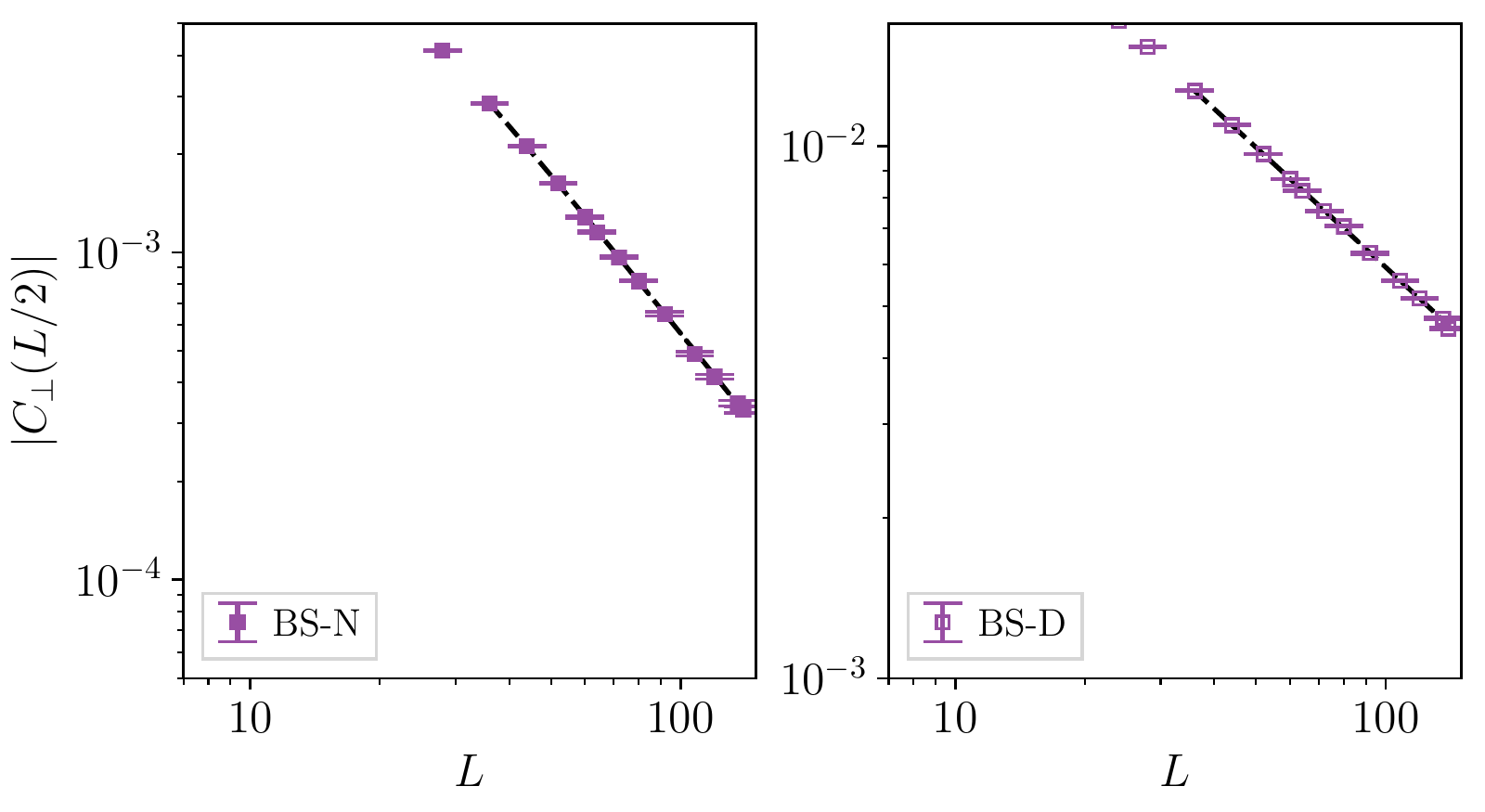} 
\includegraphics[width=\columnwidth]{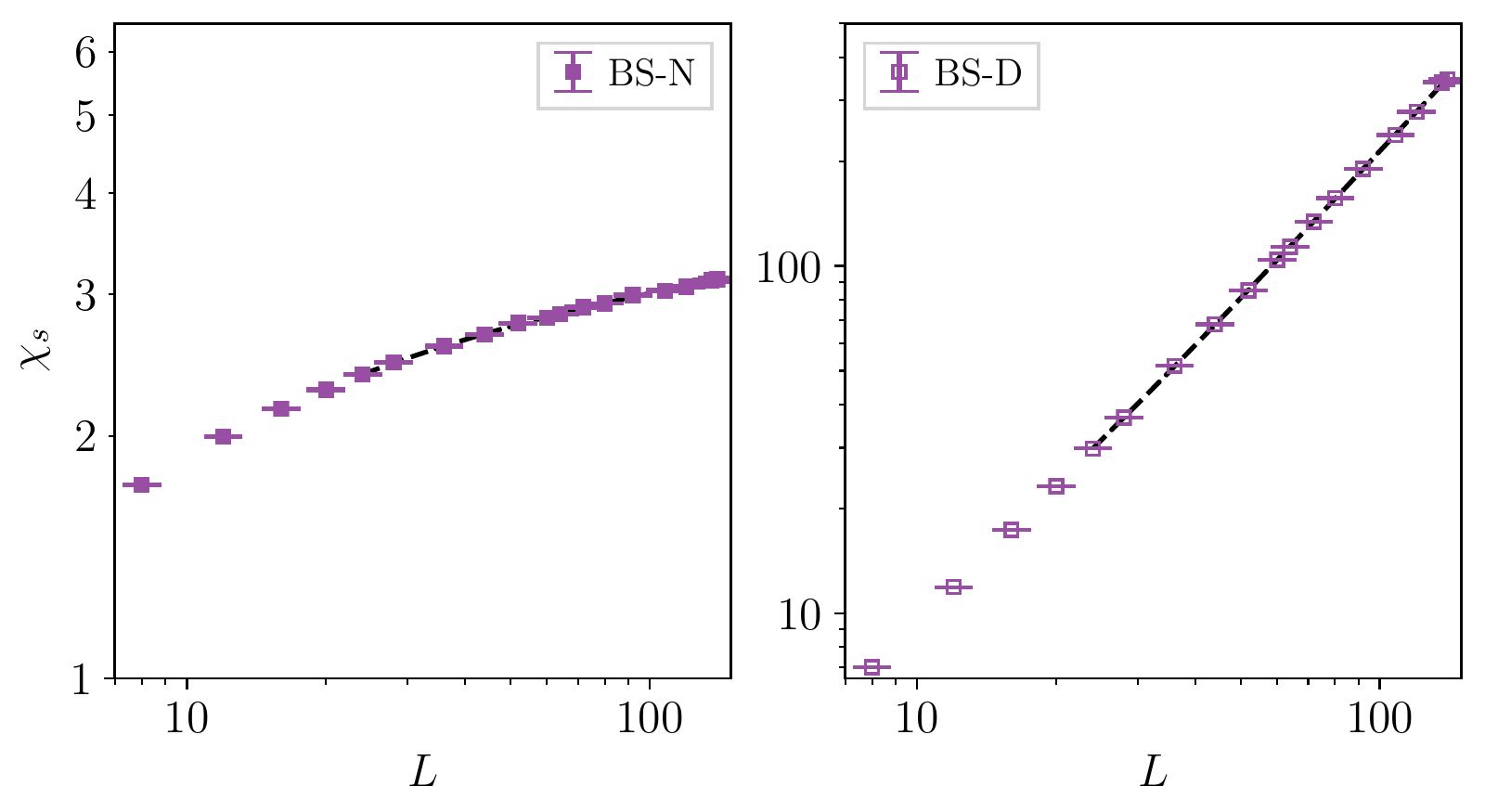} 
\end{center}
\caption{$C_\parallel(L/2)$ (top panels),  $C_\perp(L/2)$  (middle panels), and $\chi_s$  (bottom panels)  as  functions of  $L$ for the BS lattice edges shown in Fig.~\ref{fig_bilayer} along with finite-size fits (dashed lines).}
\label{fig_bilayer_scaling}
\end{figure}

In addition to the lattices discussed in the main text, we also investigate the bilayer square (BS) lattice. Here again, BS-D denotes the dangling and BS-N the nondangling edge configurations, cf.~Fig.~\ref{fig_bilayer}.
The quantum critical point for the bilayer bulk system has previously been located at $J/J_D=0.39651(2)$~\cite{Wang06}. The finite-size scaling of the QMC data for the lateral correlations $C_\parallel(L/2)$, the transverse correlatons $C_\perp(L/2)$, and the staggered susceptibility $\chi_s$ are given in Fig.~\ref{fig_bilayer_scaling}.
The critical exponents obtained from a fit to this data are given in Tab.~\ref{table3} and are consistent with the findings for the dangling and nondangling edge spin configurations from the main text.

\begingroup
\squeezetable
\begin{table}
\begin{ruledtabular}
\begin{tabular}{l || c | c | c}
Configuration & $\eta_\parallel$ & $\eta_\perp$ & $y_{h_1 }$ \\
\hline\hline
BS-N & 1.32(8) & 0.69(3) & 0.87(2)\\
\hline
BS-D & -0.49(2) & -0.25(1) & 1.733(3)\\
\end{tabular}
\end{ruledtabular}
\caption{Critical exponents $\eta_\parallel$,  $\eta_\perp$, and $y_{h_1 }$ for the edge spin configurations of the bilayer square (BS) lattice in Fig.~\ref{fig_bilayer}.
}
\label{table3}
\end{table}
\endgroup

\newpage
\subsection{Finite-size scaling analysis and scaling plots}

Figure~3 of the main tex shows the QMC data for the lateral correlations $C_\parallel(L/2)$, the transverse  correlations  $C_\perp(L/2)$, and the staggered susceptibility  $\chi_s$,  as functions of  $L$ for the different edge spin configurations shown in Fig.~1 of the main text, along with fits corresponding to the
finite-size analysis, based on the finite-size scaling ansatz given in main text. In the previous section of this supplemental material, we also present the corresponding QMC data for the BS lattice in Fig.~\ref{fig_bilayer_scaling} and the resulting exponents in Tab.~\ref{table3}.

Details concerning the range of system sizes from $L_\mathrm{min}$ to $L_\mathrm{max}$ that was accessible for the fitting procedure and whether inclusion of a scaling correction $c_1 L^{-1}$ and a nonsingular contribution $c_{ns}$ was required, are provided for each specific case in Tab.~\ref{table2}. 
In particular, a nonsingular contribution $c_{ns}$  is required for extracting $y_{h_1}$ from $\chi_s$ in the nondangling cases,
because the exponent of $\chi_s$ is negative and thus the background is the dominating term, in contrast 
to the dangling cases, where $\chi_s$ diverges and the background term would be a sub-leading correction, compared to the leading scaling correction $\propto L^{-1}$.
The $L^{-1}$ correction term was included whenever a truncation of the interval from varying $L_\mathrm{min}$  did not allow  to compatibly fit  the data  to a simple power law.
Also provided in Tab.~\ref{table2} is the formula $N(L)$  for the number of lattice sites as a function of $L$ for the various lattices.  


\begingroup
\squeezetable
\begin{table}[H]
\begin{ruledtabular}
\begin{tabular}{rlccccc}
Exponent & Config. & $L_\mathrm{min}$ & $L_\mathrm{max}$ & $N(L)$ & $c_1L^{-1}$ incl. & $c_{ns}$ incl.\\
\hline\hline
$\eta_\parallel$
& CD-D  & 30 & 140 & $L^2$   & Yes & No \\
& BS-D  & 36 & 140 & $L^2$   & Yes & No \\
& DAF-D & 18 & 88  & $4 L^2$ & Yes & No \\
& PAF-D & 18 & 88  & $4 L^2$ & Yes & No \\
& CD-N  & 30 & 140 & $L^2$   & No  & No \\
& BS-N  & 36 & 140 & $L^2$   & No  & No \\
& DAF-N & 18 & 88  & $4 L^2$ & No  & No \\
& PAF-N & 18 & 88  & $4 L^2$ & No  & No \\
\hline
$\eta_\perp$
& CD-D  & 38 & 140 & $L^2$   & No  & No \\
& BS-D  & 36 & 140 & $L^2$   & Yes & No \\
& DAF-D & 18 & 88  & $4 L^2$ & No  & No \\
& PAF-D & 18 & 80  & $4 L^2$ & No  & No \\
& CD-N  & 38 & 140 & $L^2$   & No  & No \\
& BS-N  & 36 & 140 & $L^2$   & Yes & No \\
& DAF-N & 18 & 80  & $4 L^2$ & No  & No \\
& PAF-N & 18 & 88  & $4 L^2$ & No  & No \\
\hline
$y_{h_1}$
& CD-D  & 22 & 140 & $L^2$   & Yes & No \\
& BS-D  & 24 & 140 & $L^2$   & Yes & No \\
& DAF-D & 14 & 88  & $4 L^2$ & Yes & No \\
& PAF-D & 14 & 88  & $4 L^2$ & Yes & No \\
& CD-N  & 22 & 140 & $L^2$   & No  & Yes \\
& BS-N  & 24 & 140 & $L^2$   & No  & Yes \\
& DAF-N & 14 & 88  & $4 L^2$ & No  & Yes \\
& PAF-N & 14 & 88  & $4 L^2$ & No  & Yes \\
\end{tabular}
\end{ruledtabular}
\caption{Details of the fitting range and fitting formula for the critical exponents $\eta_\parallel$,  $\eta_\perp$, and $y_{h_1 }$ for the different edge spin configurations of 2D coupled spin-dimer systems. 
}
\label{table2}
\end{table}
\endgroup

\newpage

\begin{figure}[H]
\begin{center}
\includegraphics[width=\columnwidth]{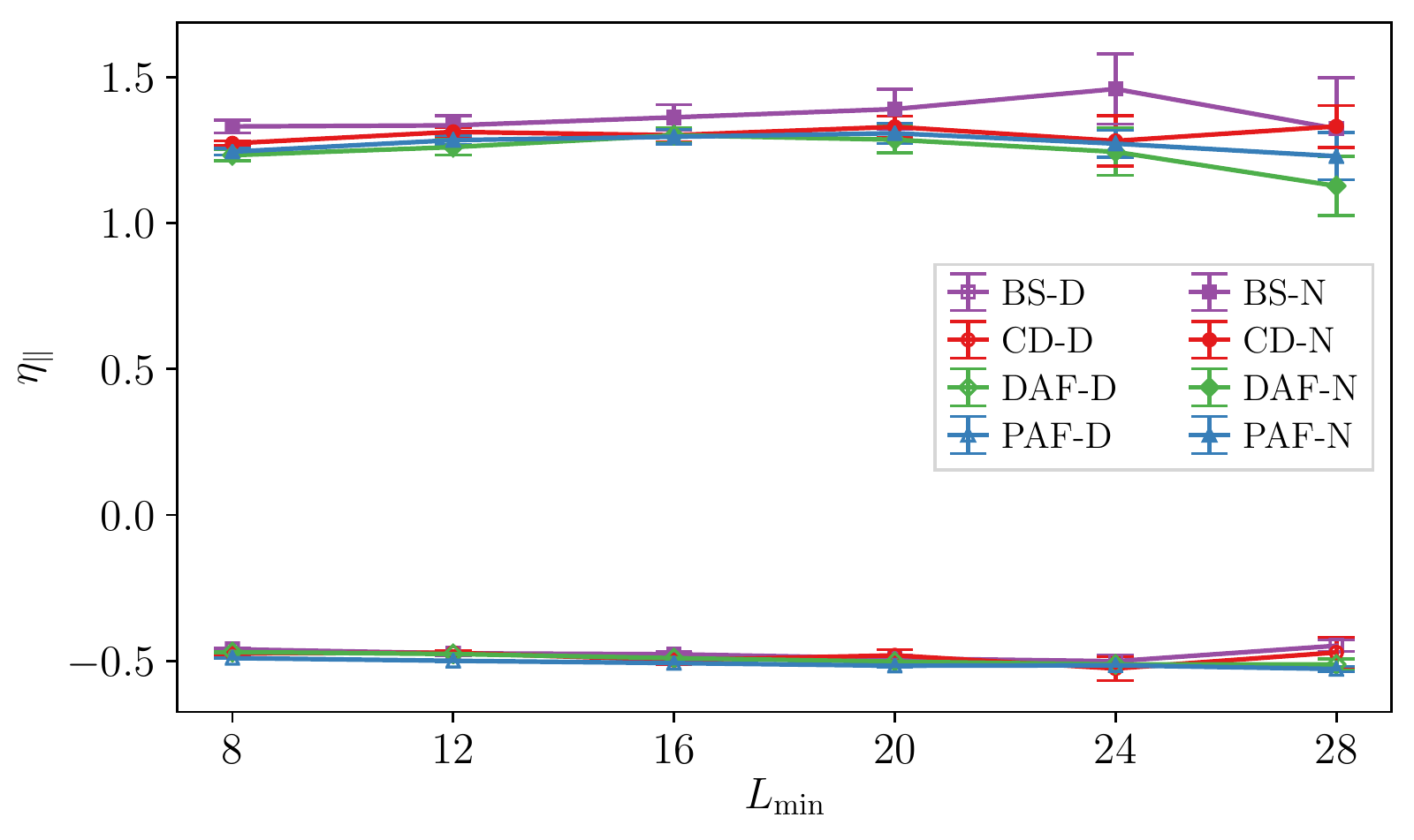} 
\includegraphics[width=\columnwidth]{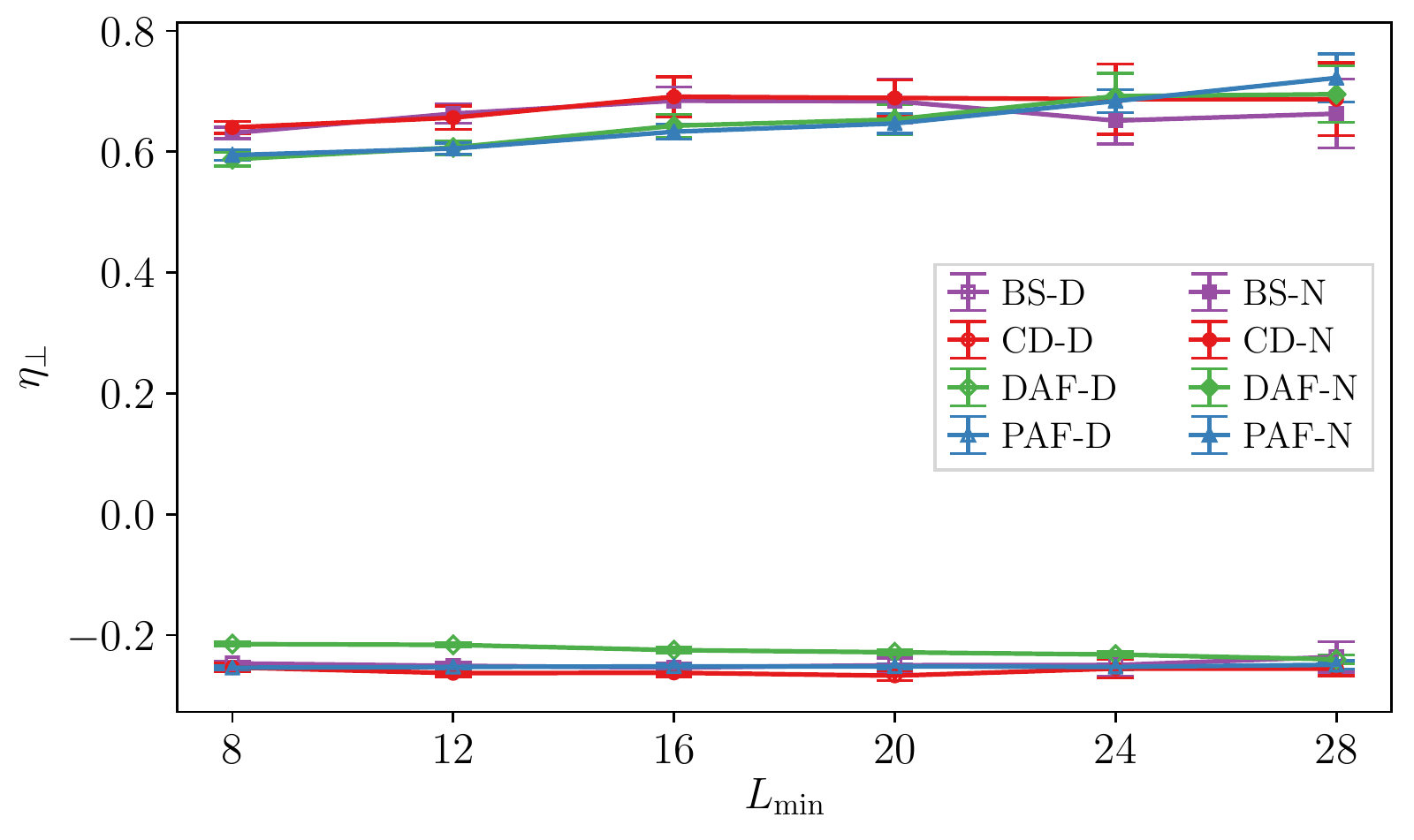} 
\includegraphics[width=\columnwidth]{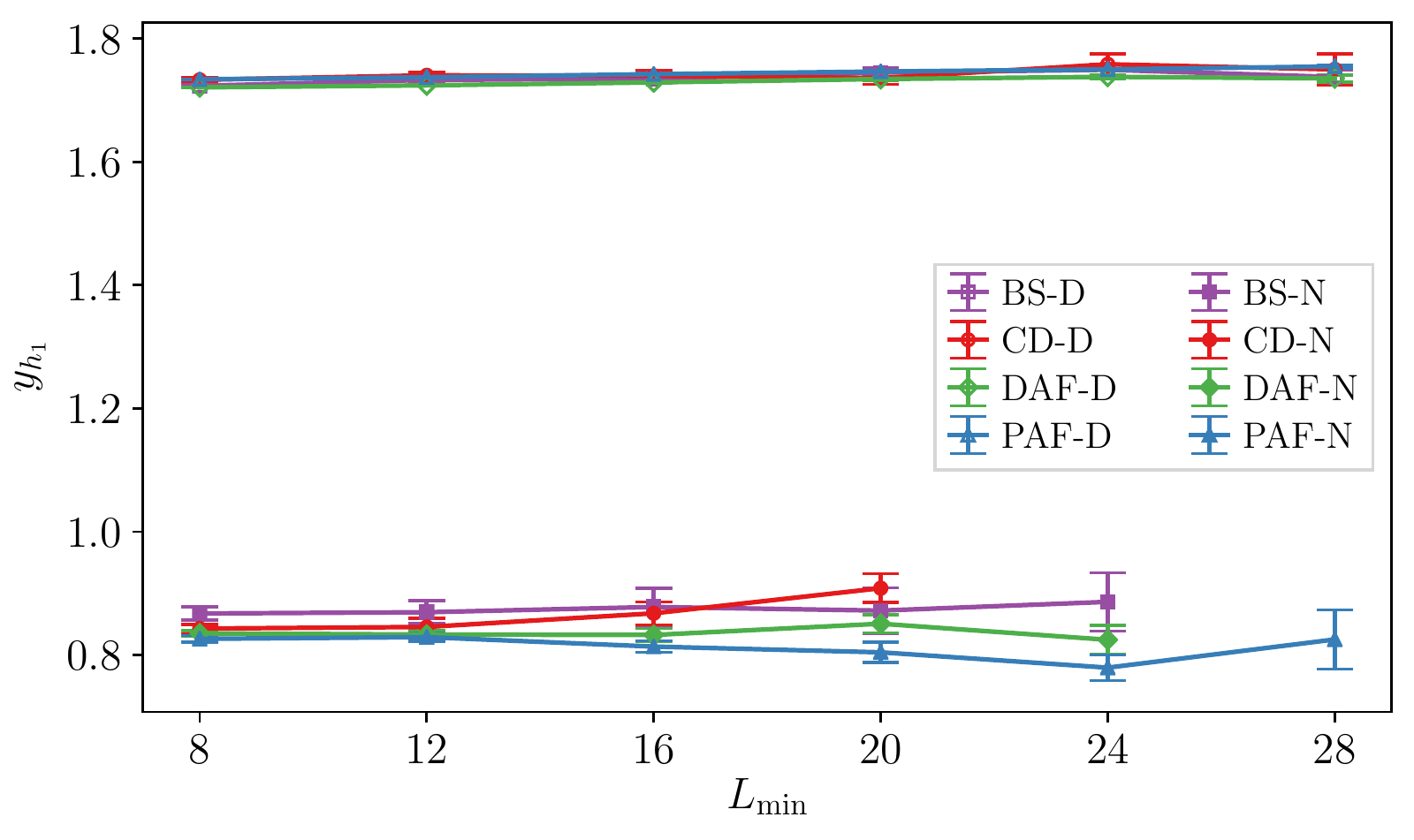} 
\end{center}
\caption{Dependence of the estimate for $\eta_\parallel$  (top panel), $\eta_\perp$  (middle panel), and $y_{h_1}$ (bottom panel)  on the value of $L_\mathrm{min}$ for the various considered edge spin configurations. }
\label{fig_lmin1}
\end{figure}

We furthermore monitored the dependence of the extracted exponents on the minimum lattice size $L_\mathrm{min}$ included in the fitting procedure. The dependence of the various critical exponents on $L_\mathrm{min}$ is shown in Fig.~\ref{fig_lmin1}. We observe no signifiant $L_\mathrm{min}$-dependence apart from strongly increasing uncertainties for the larger values of  $L_\mathrm{min}$, reflecting the fact that  data from fewer system sizes are then available for the fitting procedure. 

\begin{figure}[H]
\begin{center}
\includegraphics[width=\columnwidth]{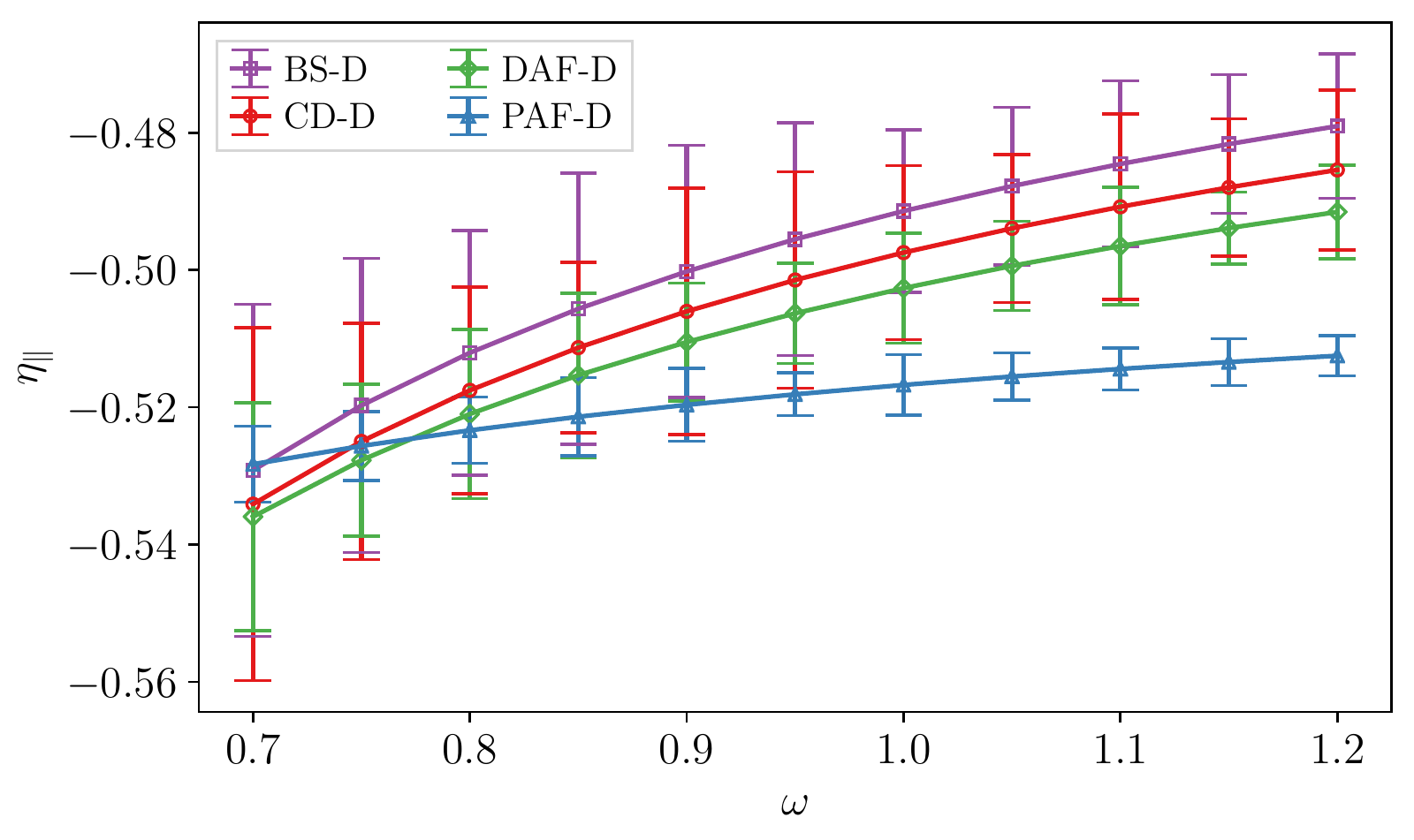} 
\includegraphics[width=\columnwidth]{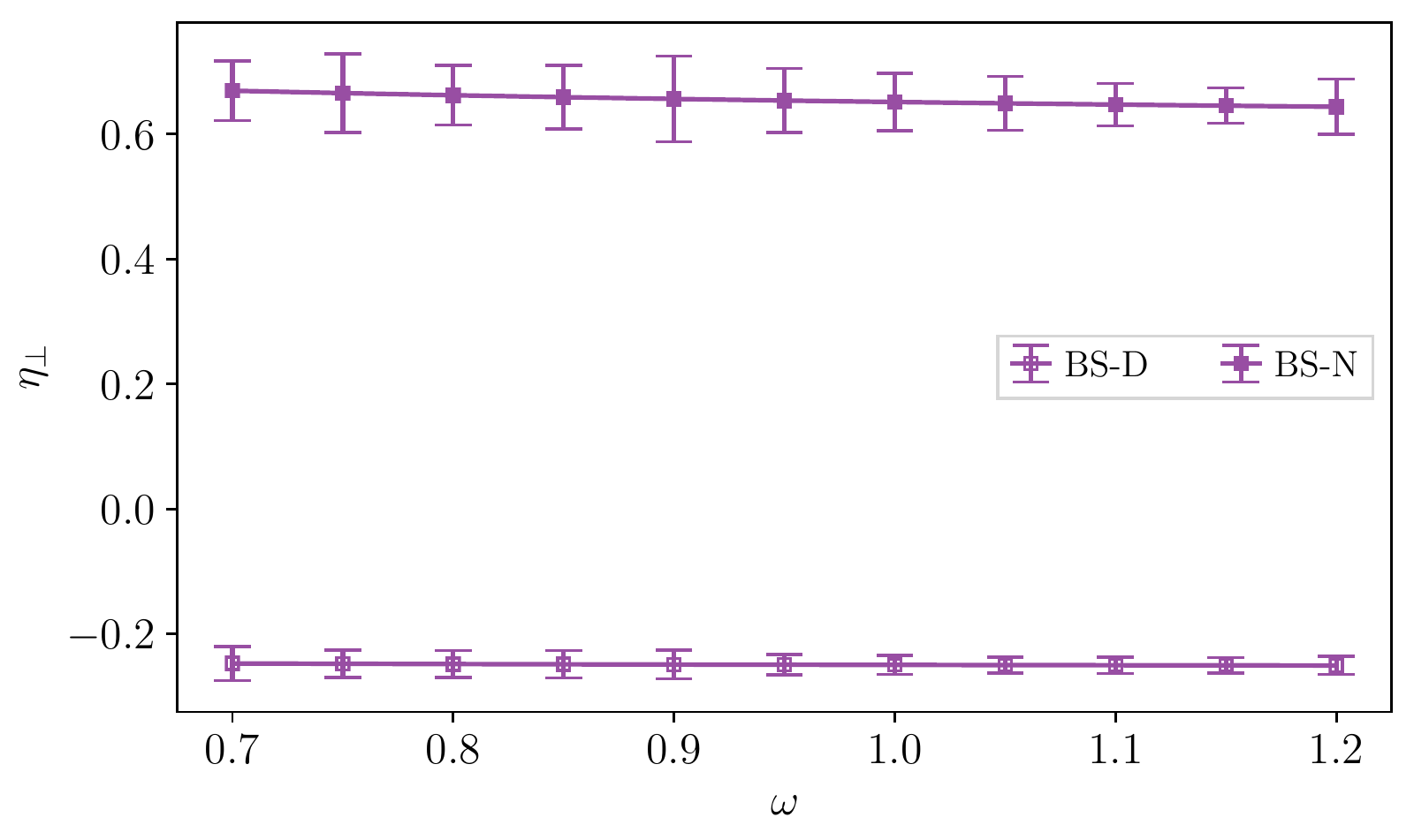} 
\includegraphics[width=\columnwidth]{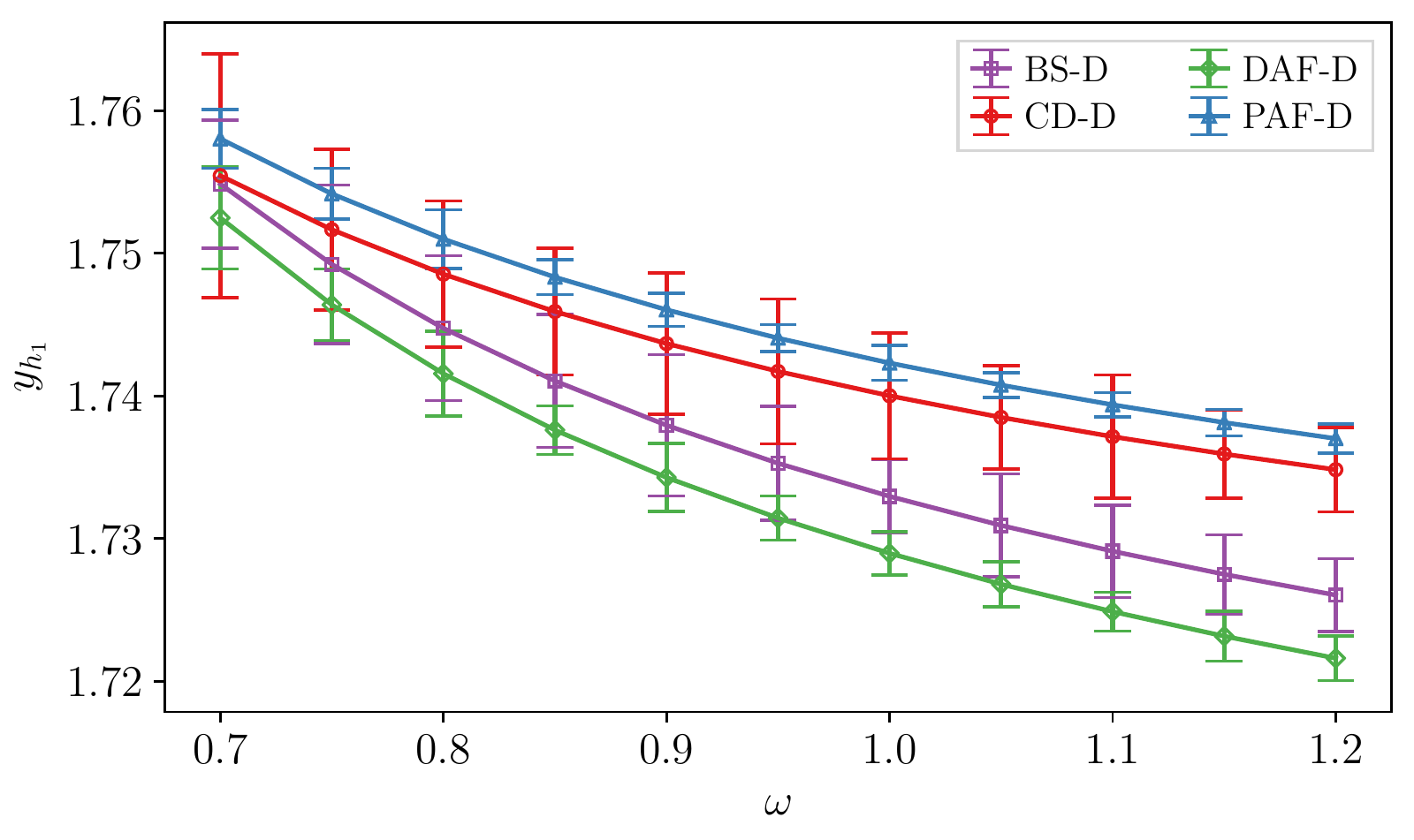} 
\end{center}
\caption{Dependence of the estimate for $\eta_\parallel$ (top panel), $\eta_\perp$  (middle panel), and  $y_{h_1}$ (bottom panel) on the value of $\omega$ for the various considered edge spin configurations.  }
\label{fig_w1}
\end{figure}

Finally, we  also considered a  finite-size scaling ansatz, wherein the  scaling correction $\propto L^{-1}$  is replaced by a more general form $\propto L^{-\omega}$, with a free exponent $\omega$. For example,
 $\omega \approx 0.8$  corresponds to the correction-to-scaling exponent of the classical O(3) model at the 3D bulk phase transition~[34]. As shown in Fig.~\ref{fig_w1}, we observe only mild trends in the $\omega$-dependence for some of the exponents. Anticipating the statistical uncertainties on the accessible system sizes,  no significant qualitative changes result for the estimated exponents. Based on the above considerations, we thus consider the exponents given in  Tab.~I of the main text to provide reliable estimates for the current purpose of distinguishing the two different cases of dangling vs nondangling edge spin configurations. 
 
\newpage

\subsection{Connecting the plaquette-square and  bilayer  models}

\begin{figure}[H]
\begin{center}\vspace{1cm}
\includegraphics[width=0.8\columnwidth]{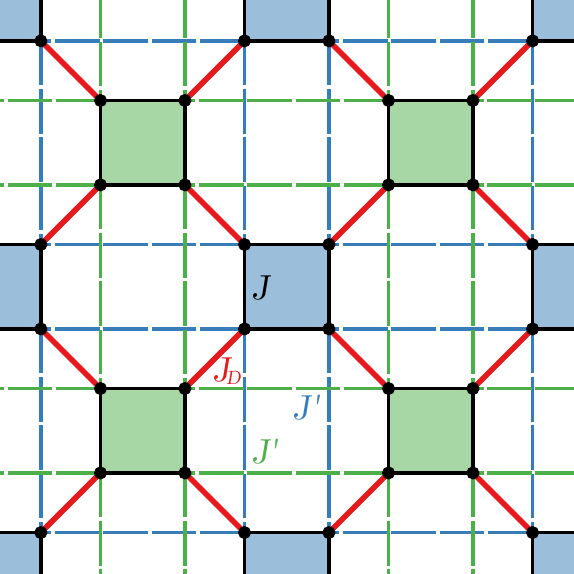} 
\end{center}
\caption{Illustration showing how the bilayer square lattice is obtained from the  plaquette-square  lattice model upon increasing the couplings $J'$, indicated by dashed lines, from $0$ to the value of  $J$.
}
\label{fig_psb}
\end{figure}

Figure~\ref{fig_psb} illustrates, how the bilayer lattice model  is  obtained from the plaquette-square lattice model upon increasing the additional exchange couplings $J'$ from $0$ to the value of $J$. During this process neither the internal SU(2) symmetry nor the spatial symmetries of the original plaquette square lattice are broken. 
The dimer bonds $J_D$ thereby become the perpendicular inter-layer bonds. From explicit QMC calculations for different values of  $J/J_D$ inside the dimerized phase, one indeed obtains no indication for a quantum phase transition during this increase of $J'$, neither from the ground state energy nor the fidelity susceptibility. 
To relate to the more conventional
presentation of the bilayer model, one may  consider shifting all the blue (green) bonds and plaquettes  up (down) to form the  upper (lower) square lattice. 

\begin{references}
\bibitem{footnote2} Our estimated exponents for the DAF-D and PAF-N cases are consistent with those reported in 
Ref.~\cite{Zhang17}. Reference~\cite{Zhang17} also provides estimates of the critical exponents for a dangling edge spin configuration of the PS model at the quantum phase transition  between the dimerized phase and the ferromagnetic regime at $J/J_D=-0.934251(11)$. The estimated values of the critical exponents, $\eta_\parallel=-0.561(4)$, $\eta_\perp=-0.2707(24)$ and $y_{h_1}=1.7802(16)$ also differ slightly from the obtained values for the DAF-D case.
\end{references}
\end{document}